\documentclass[aps,prl,showpacs,preprintnumbers,twocolumn,groupedaddress]{revtex4}
\usepackage{graphicx}
\usepackage{amsmath}

\begin{document}

\preprint{RIKEN-QHP-52}
\title{Lattice QCD with strong external electric fields}
\author{Arata~Yamamoto}
\affiliation{Theoretical Research Division, Nishina Center, RIKEN, Saitama 351-0198, Japan}

\date{\today}

\begin{abstract}
We study particle generation by a strong electric field in lattice QCD.
To avoid the sign problem of the Minkowskian electric field, we adopt the ``isospin'' electric charge.
When a strong electric field is applied, the insulating vacuum is broken down and pairs of charged particles are produced by the Schwinger mechanism.
The competition against the color confining force is also discussed.
\end{abstract}


\maketitle

\emph{Introduction.}---
Quarks interact with not only gluons but also photons.
Because the electromagnetic interaction is much weaker than the strong interaction, the electromagnetic interaction has been neglected in most lattice QCD simulations.
However, the electromagnetic interaction can be essential in several situations \cite{Tiburzi:2011vk}.
For example, high precision simulations including dynamical QED effects are currently possible.
Such simulations can reproduce electromagnetic properties of hadrons.
Also, strong background electromagnetic fields are important for QCD phenomenology.
Even if the electromagnetic coupling is smaller than the QCD coupling, strong electromagnetic fields can drastically affect hadron properties.
A strong external magnetic field is one hot topic in lattice QCD \cite{Yamamoto:2012bi}.

Compared to the case of magnetic fields, the implementation of electric fields is not simple.
Two kinds of electric fields are possible: the Euclidean electric field and the Minkowskian electric field.
(In the following discussion, we choose the axial gauge $A_j(x)=0$.)

For the Euclidean electric field, the Euclidean gauge field $A_4(x)$ is introduced as
\begin{equation}
i \partial_0 +qA_0(x)-\mu \to -\partial_4 +iqA_4(x) -\mu
\end{equation}
in the Wick rotation.
This gauge field $A_4(x)$ respects the U(1) symmetry in the Euclidean space.
As understood from the relation to the chemical potential $\mu$, the Euclidean electric field cannot describe particle generation.
The Euclidean electric field has been used for calculating electric polarizabilities in lattice simulations \cite{Fiebig:1988en}.
The analytic continuation is needed to obtain the physical result in the Minkowski space.

On the other hand, for the Minkowskian electric field, the vector potential $A_0(x)$ is independent of the transformation
\begin{equation}
i \partial_0 +qA_0(x)-\mu \to -\partial_4 +qA_0(x) -\mu.
\label{eqAM}
\end{equation}
This electric field is a real electric field in the sense that the generated electric energy is real (not imaginary).
The implementation of the Minkowskian electric field is difficult in lattice simulations.
The vector potential breaks the anti-Hermitian property of the Dirac operator, and thus causes the notorious sign problem.
The Minkowskian electric field has been applied only in (partially) quenched lattice QCD \cite{Shintani:2006xr}.
Moreover, a constant Minkowskian electric field is impossible in a periodic box because the vector potential is a real number, as explained later.

\emph{Setups.}---
The sign problem of the Minkowskian electric field is completely the same as that of a chemical potential.
As seen in Eq.~(\ref{eqAM}), the vector potential $A_0(x)$ is regarded as a coordinate-dependent chemical potential.
This means that the same strategy is applicable to the Minkowskian electric field and a chemical potential.
The sign problem occurs in the physical electric charge $q = {\rm diag} (q_u,q_d) = {\rm diag} (2e/3,-e/3)$.
To avoid the sign problem, we adopt a specific choice of the electric charges,
\begin{equation}
q_3 \equiv e \frac{\sigma_3}{2} = {\rm diag} \left( +\frac{e}{2}, -\frac{e}{2} \right),
\end{equation}
which we call the ``isospin'' electric charge.
This prescription is inspired by an isospin chemical potential $\mu_3 \equiv \mu \sigma_3/2 $, which does not cause the sign problem \cite{Kogut:2002zg}.
The complex phases of the fermion determinants are canceled out between the $u$-quark and $d$-quark sectors.
Thus, the fermion determinant is semi-positive and the sign problem does not occur.

\begin{figure}[h]
\begin{center}
\includegraphics[scale=0.5]{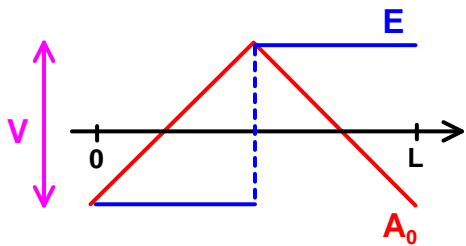}
\caption{\label{fig1}
The configuration of the vector potential $A_0(z)$ and the electric field $E(z)$.
The boundary condition is periodic.
}
\end{center}
\end{figure}

For numerical simulations, we consider the external Minkowskian electric field in a finite box with periodic boundary conditions.
To satisfy the periodic boundary condition, we divide a finite spatial size $L$ to two regions; $L > z \ge L/2$ and $L/2 > z \ge 0$.
We set the vector potential as
\begin{equation}
A_0(z) = 
\begin{cases}
+E_0(z-\frac{L}{4}) &(\frac{L}{2} > z \ge 0)\\
-E_0(z-\frac{3L}{4}) &(L > z \ge \frac{L}{2}).
\end{cases}
\label{eqA0}
\end{equation}
All the spatial components are zero, $A_j(x)=0$. 
This configuration is depicted in Fig.~\ref{fig1}.
In this configuration, the electric field is applied in the $z$-direction,
\begin{equation}
E(z) = - \partial_z A_0(z) =
\begin{cases}
-E_0 &(\frac{L}{2} > z \ge 0)\\
+E_0 &(L > z \ge \frac{L}{2}).
\end{cases}
\end{equation}
The voltage difference is $V = E_0L/2$.

On the lattice, the vector potential is introduced as real link variables $u_0(x) =e^{q_3A_0(x)}$ and $u_0^{-1}(x) =e^{-q_3A_0(x)}$.
For example, the Wilson Dirac operator with the vector potential is
\begin{eqnarray}
D &=& 1
- \kappa \sum_i \bigl[ (1-\gamma_i)T_{i+} \nonumber + (1+\gamma_i)T_{i-} \bigr] \nonumber\\
&&- \kappa \bigl[ (1-\gamma_4)u_0^{-1}T_{4+} + (1+\gamma_4)u_0T_{4-} \bigr] 
\end{eqnarray}
with $[T_{\mu +}]_{x,y} \equiv U_\mu (x) \delta_{x+\hat{\mu},y}$ and
$[T_{\mu -}]_{x,y} \equiv U^\dagger_\mu (y) \delta_{x-\hat{\mu},y}$.
This is the same implementation as an isospin chemical potential \cite{Kogut:2002zg,Hasenfratz:1983ba}.

Since the link variable $u_\mu(x)$ is not a U(1) element, the U(1) gauge symmetry is lost in the Euclidean space.
However, there is a corresponding gauge symmetry.
Let us consider the transformation
\begin{eqnarray}
A_0(x) &\to& A_0(x) + \partial_4 \Lambda(x) \\
A_j(x) &\to& A_j(x) -i \partial_j \Lambda(x).
\end{eqnarray}
This transformation corresponds to the U(1) gauge transformation in the Minkowski space by replacing $\Lambda(x) \to i\Lambda(x)$.
The Minkowskian electric field
\begin{equation}
E_j(x) = i\partial_4 A_j(x) - \partial_j A_0(x)
\end{equation}
is invariant under this transformation.
Note that $A_j(x)$ must be a pure imaginary number.
On the lattice, the corresponding transformation is
\begin{eqnarray}
u_0(x) =e^{q_3A_0(x)} &\to& e^{-q_3\Lambda(x)} u_0(x) e^{q_3\Lambda(x+\hat{4})} \\
u_j(x) =e^{iq_3A_j(x)} &\to& e^{-q_3\Lambda(x)} u_j(x) e^{q_3\Lambda(x+\hat{j})}.
\end{eqnarray}
By the same argument as the gauge invariance on the lattice, we can prove that any gauge-invariant operator, i.e., a closed loop of the link variables, is invariant under this transformation.

\emph{Physical picture.}---
What we expect in the Minkowskian electric field is as follows.
When a strong electric field is applied, quark-antiquark pairs are produced from the vacuum by the Schwinger mechanism \cite{Schwinger:1951nm}.
The quarks and the antiquarks flow along the electric field.
However, we cannot observe these nonequilibrium processes.
In lattice QCD simulations, the observable must be in an equilibrium state.
The equilibrium state in the electric field is the final state after the charged particles flow and stop.
In a finite box, the highest and lowest voltage regions exist somewhere.
Positive charged particles stop at the lowest voltage region and negative charged particles stop at the highest voltage region.
As a consequence, a nonuniform charge density distribution appears.
When the voltage difference increases, the charge density grows. 

\begin{figure}[h]
\begin{center}
\includegraphics[scale=0.45]{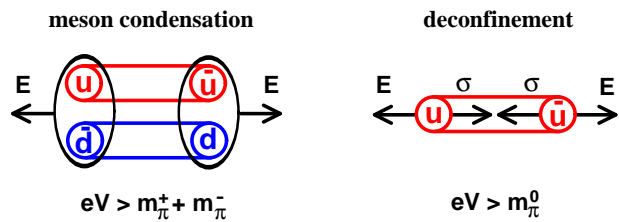}
\caption{\label{fig2}
The meson condensation and the deconfinement.
}
\end{center}
\end{figure}

In the confinement phase, we need to take into account color confinement.
Charged particles are created by the Schwinger mechanism, but they cannot flow freely due to the confining force.
There are two possibilities to separate the charged particles in the confinement phase.
The schematic figure is shown in Fig.~\ref{fig2}.
The first possibility is meson condensation.
Charged mesons can be formed when the voltage difference exceeds twice the lightest charged meson mass.
The charge generation occurs only above this threshold.
Because the lightest charged meson is a pion, the threshold is twice the charged pion mass.
This is similar to the charged pion condensation at a finite isospin density, where the charged pion is condensed in $\mu_3 \ge m_\pi$ \cite{Son:2000xc}.
The second possibility is the deconfinement.
The electric field draws the charged particles.
This force is oriented in the opposite direction to the color confining force.
When the electric field overcomes the confining force, the charged particles can be separated.
The threshold of the voltage is the neutral pion mass.
This effect is sensitive to the system volume.
When the quark mass is small, the spatial size of the quark-antiquark pair gradually increases.
Thus, the charge density gradually appears in a finite volume even if the electric field is smaller than the confining force, while this effect is suppressed in a larger volume.
Note that, quantitatively, these thresholds can be shifted from the original pion mass by the effect of the electric field.

If we can apply a constant electric field in a periodic box, we can generate a permanent electric current.
In the case of the Minkowskian electric field, we cannot generate a  permanent electric current.
In periodic boundary conditions, a constant voltage gradient is impossible and only the configuration like Fig.~\ref{fig1} is possible.
We cannot make a circuit with a steady electric current.
(In the case of the Euclidean electric field or a magnetic field, a constant field is possible in a periodic box \cite{AlHashimi:2008hr}.
This is because the U(1) link variable is a complex phase factor which has $2\pi$ periodicity.
Actually, a constant vector current was measured in the lattice simulation with a magnetic field \cite{Yamamoto:2011gk}.)

\emph{Simulation.}---
We performed the two-flavor full QCD simulation with the plaquette gauge action and the Wilson fermion action.
The vector potential is included both in the valence and dynamical fermions.
The gauge coupling is $\beta = 5.32441$ and the hopping parameter is $\kappa = 0.1665$.
The lattice spacing is $a\simeq 0.13$ fm and the pion mass is $m_\pi \simeq 0.4$ GeV \cite{Orth:2005kq}.
The spatial lattice volume is $L^3 = (12 a)^3$ with periodic boundary conditions.
We used two temporal lattice sizes; $N_\tau=12$ for the confinement phase and $N_\tau=4$ for the deconfinement phase.

\begin{figure}[h]
\begin{center}
\includegraphics[scale=1]{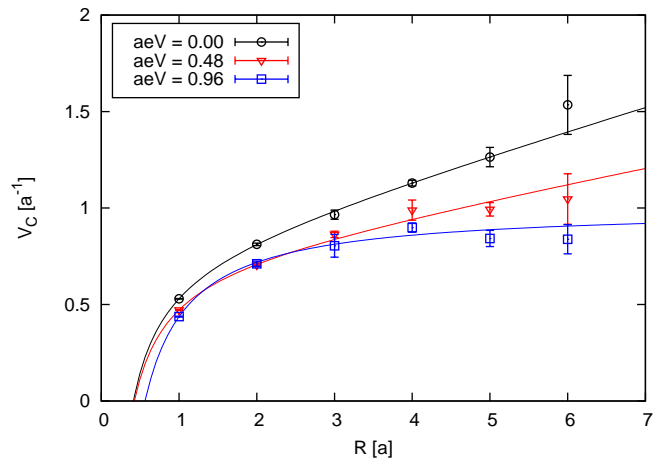}
\caption{\label{fig3}
The charged heavy-quark potential $V_C(R)$.
}
\end{center}
\end{figure}

First, we demonstrate the competition between the electric field and the color confining force.
We consider a ``charged'' heavy-quark potential.
The heavy-quark potential is extracted from the Wilson loop.
When the heavy quarks have electric charges, the link variable $u_\mu(x)$ is multiplied to the SU(3) link variable $U_\mu(x)$ as
\begin{equation}
W_C \equiv {\rm tr} \prod_{\rm loop} \{ U_\mu(x)u_\mu(x) \} = W_{SU(3)} \prod_{\rm loop} u_\mu(x).
\end{equation}
For the charged heavy-quark potential in a constant electric field, we calculate the rectangular ($R\times T$) Wilson loop in the $z$-$t$ plane.
From Eq.~(\ref{eqA0}), the expectation value of the Wilson loop is
\begin{equation}
\langle W_C(R,T) \rangle = \langle W_{SU(3)}(R,T) \rangle e^{\frac{e}{2}E_0RT},
\end{equation}
and the charged heavy-quark potential is
\begin{equation}
V_C(R) =  V_{SU(3)}(R) - \frac{e}{2}E_0R.
\end{equation}
The sign of the second term depends on the electric charge and the region ($L > z \ge L/2$ or $L/2 > z \ge 0$).
We here choose the repulsive case.
In quenched QCD, the SU(3) heavy-quark potential $V_{SU(3)}(R)$ is independent of the electric field, and it is the so-called Cornell potential.
Thus,
\begin{eqnarray}
V_C(R) =  \left( \sigma - \frac{e}{2}E_0 \right) R +\frac{A}{R} + {\rm const} .
\label{eqVC}
\end{eqnarray}
The electric field suppresses the linear confining potential.
In full QCD, $V_{SU(3)}(R)$ can be modified by the electric field through the dynamical quark.
In Fig.~\ref{fig3}, we show the numerical result of the charged heavy-quark potential.
The data are fitted by Eq.~(\ref{eqVC}).
Although this is the full QCD simulation, the charged heavy-quark potential is consistent with the quenched QCD form (\ref{eqVC}).
At $aeV = 0.96$, the linear confining potential disappears because the electric field is almost the same as the string tension, $a^2eE_0/2 = 0.08 \simeq a^2\sigma$.

\begin{figure}[h]
\begin{center}
\includegraphics[scale=1]{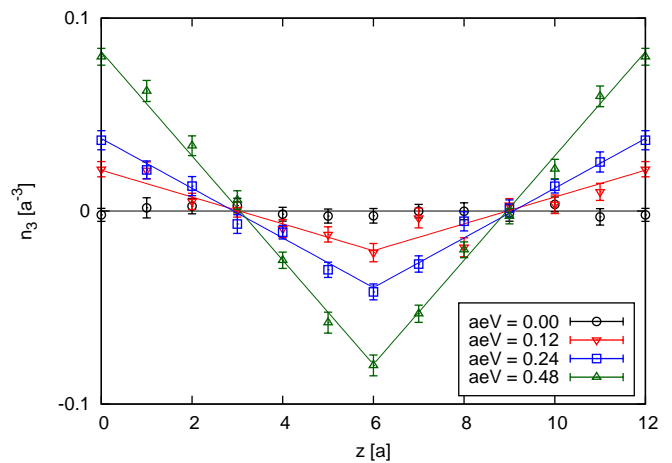}
\caption{\label{fig4}
The charge density distribution $n_3(z)$.
}
\end{center}
\end{figure}

Next, we calculated the charge density
\begin{equation}
n_3(x) \equiv \frac{1}{e} \frac{\partial \ln Z}{\partial A_0(x)} = \frac{1}{e} \left\langle{\rm Tr} D^{-1} \frac{\partial D}{\partial A_0(x)} \right\rangle.
\end{equation}
This charge density is equivalent to the isospin density \cite{Kogut:2002zg}.
In Fig.~\ref{fig4}, we show the charge density distribution in the deconfinement phase.
The charge density was fitted by a combination of two linear functions.
When the voltage is applied, the charge density becomes finite in the $A_0 \ne 0$ region.
The positive (negative) charge density means the appearance of the positive (negative) charged particles, i.e., $u$-quarks and $\bar{d}$-quarks ($\bar{u}$-quarks and $d$-quarks).
The charge density increases when the voltage increases.

\begin{figure}[h]
\begin{center}
\includegraphics[scale=1]{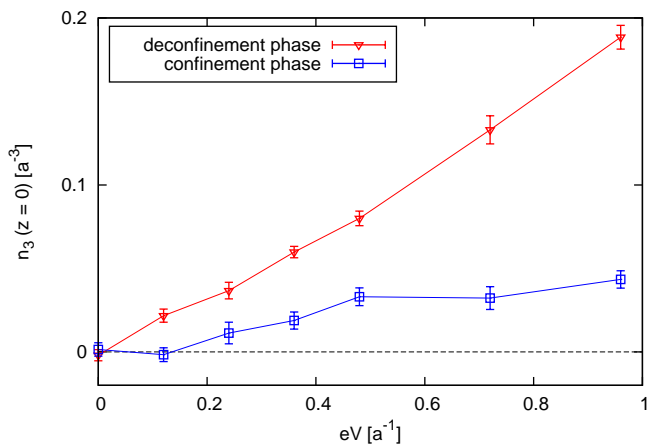}
\caption{\label{fig5}
The voltage dependence of the charge density $n_3$ at $z=0$.
}
\end{center}
\end{figure}

In Fig.~\ref{fig5}, we show the voltage dependence of the charge density at $z=0$.
We calculated at the voltage $aeV=0$, 0.12, 0.24, 0.36, 0.48, 0.72, and 0.96.
We show the data in the deconfinement phase and in the confinement phase, where only the temporal lattice size $N_\tau$ is changed and other parameters ($\beta$, $\kappa$, and $L$) are fixed at the values explained above.
In the deconfinement phase, the charge density grows monotonically because the charged particles flow freely. 
In the confinement phase, the interpretation of the result is nontrivial.
If the charge density were generated by the pion condensation, the charge density would be zero in $eV < 2m_\pi$.
The pion mass at $V=0$ is $am_\pi \simeq$ 0.26 \cite{Orth:2005kq}.
The charge density is finite in $aeV < 2am_\pi \simeq 0.52$.
The charge density in $eV < 2m_\pi$ is generated not by the meson condensation but by the deconfinement in a finite volume.
The charge density is finite even at $aeV=0.24$, which is slightly below $m_\pi$. 
This is not surprising because the pion mass will be shifted by the electric field.
In a smaller voltage $eV \ll m_\pi$, the charge density is not generated.
This is consistent with the expectation that the QCD vacuum is an insulator at zero temperature.

To know the asymptotic behavior, we need to take the large system limit $L \to \infty$.
We can consider two types of the large system limit.
In the $L \to \infty$ limit with fixed $V$ (i.e., $E \to 0$), the electric field cannot induce the deconfinement.
The charge density appears only in $eV \ge 2m_\pi$.
On the other hand, in the $L \to \infty$ limit with fixed $E$ (i.e., $V \to \infty$), the system energy diverges.
An infinite number of charged mesons are produced from the vacuum.
The deconfinement occurs only in $eE/2 \ge \sigma$.

\emph{Summary.}---
We have simulated the Minkowskian electric field by introducing the isospin electric charge.
Similarly, we can adopt two-color QCD, which can also avoid the sign problem.
In these frameworks, it is possible to study QCD in electric fields.
We have analyzed the charge particle generation by the Schwinger mechanism.
Although we cannot observe the nonequilibrium process itself, we can study some nonperturbative aspects of the particle generation.
The particle generation is characteristic in the Minkowskian electric field because the vacuum is stable in the Euclidean electric field.

The author greatly thanks Yoshimasa Hidaka for useful discussions.
The author is supported by the Special Postdoctoral Research Program of RIKEN.
The lattice QCD simulations were carried out on NEC SX-8R in Osaka University.

\end{document}